\documentclass[12pt,preprint]{aastex}

\newcommand{\Mo}{$M_{\odot}$}
\newcommand{\MJ}{$M_{J}$}
\newcommand{\ms}{m s$^{-1}$}

\shorttitle{On the Mass-Period Correlation of Extrasolar Planets}
\shortauthors{Zucker \& Mazeh}

\begin{document}

\title{On the Mass-Period Correlation of the Extrasolar Planets}

\author{Shay Zucker and Tsevi Mazeh}
\affil{School of Physics and Astronomy, Raymond and Beverly Sackler
Faculty of Exact Sciences, Tel Aviv University, Tel Aviv, Israel}
\email{shay@wise.tau.ac.il; mazeh@wise.tau.ac.il} 

\begin{abstract}

We report on a possible correlation between the masses and periods
of the extrasolar planets, manifested as a paucity of massive planets
with short orbital periods. Monte-Carlo simulations show the
effect is significant, and is not solely due to an
observational selection effect. We also show the effect is
stronger than the one already implied by published models that assumed
independent power-law distributions for the masses and periods of the
extrasolar planets.  Planets found in binary stellar systems may have
an opposite correlation. The difference is highly significant despite
the small number of planets in binary systems. We discuss the paucity
of short-period massive planets in terms of some theories for the close-in
giant planets. Almost all models can account for the deficit of
massive planets with short periods, in particular the model that
assumes migration driven by a planet-disk interaction, if the planet
masses do not scale with their disk masses.

\end{abstract}


\section{INTRODUCTION}

The mass distribution of the extrasolar planets was recognized to be a
key feature of the growing new population since the first few
detections (e.g., Basri \& Marcy 1997; Mayor, Queloz \& Udry 1998;
Mazeh, Goldberg \& Latham 1998; Heacox 1999; Mazeh 1999; Stepinski \&
Black 2000). Recent studies showed that the mass
distribution is probably flat or slightly decreasing in log~M (Jorissen, 
Mayor \& Udry 2001; Zucker \& Mazeh 2001; Tabachnik \& Tremaine 2002;
Lineweaver \& Grether 2002), and has a distinct cutoff at about 10 Jupiter 
masses(=\MJ).  This paper focuses on a possible dependence between the
extrasolar planets masses and their orbital periods.

Every study of the mass-period relation has to take into account the
strong observational selection effect that prohibits the detection of
low-mass--long-period planets, because of their small radial-velocity
amplitudes.  Tabachnik \& Tremaine (2002) studied the mass and the
period distributions simultaneously, assuming they were two
independent power-law distributions. They found that the uncertainties
of the exponents of the two variables are highly correlated, but
attributed their findings to the observational selection effect. This
paper shows that the mass-period correlation found in the sample
of known extrasolar planets {\it cannot} be attributed solely to the
observational selection effect.  We show that there is an additional
real dependency between the mass and the period of the extrasolar
planets, manifested as a significant paucity of
high-mass--short-period planets. Since such planets are the easiest to
detect, this paucity is probably not the result of any selection
effect.

In Section~2 we present our analysis of the mass-period correlation of
the whole sample of known planets. Section~3 shows that the small
subsample of planets in binary stellar systems may have different,
opposite, correlation. Section~4 shortly discusses our findings in
terms of some theories for the existence of giant planets close to
their host stars, the migration model in particular.  A preliminary
version of this work was presented in Mazeh \& Zucker (2002).

\section{Analysis}

\begin{figure}
\plotone{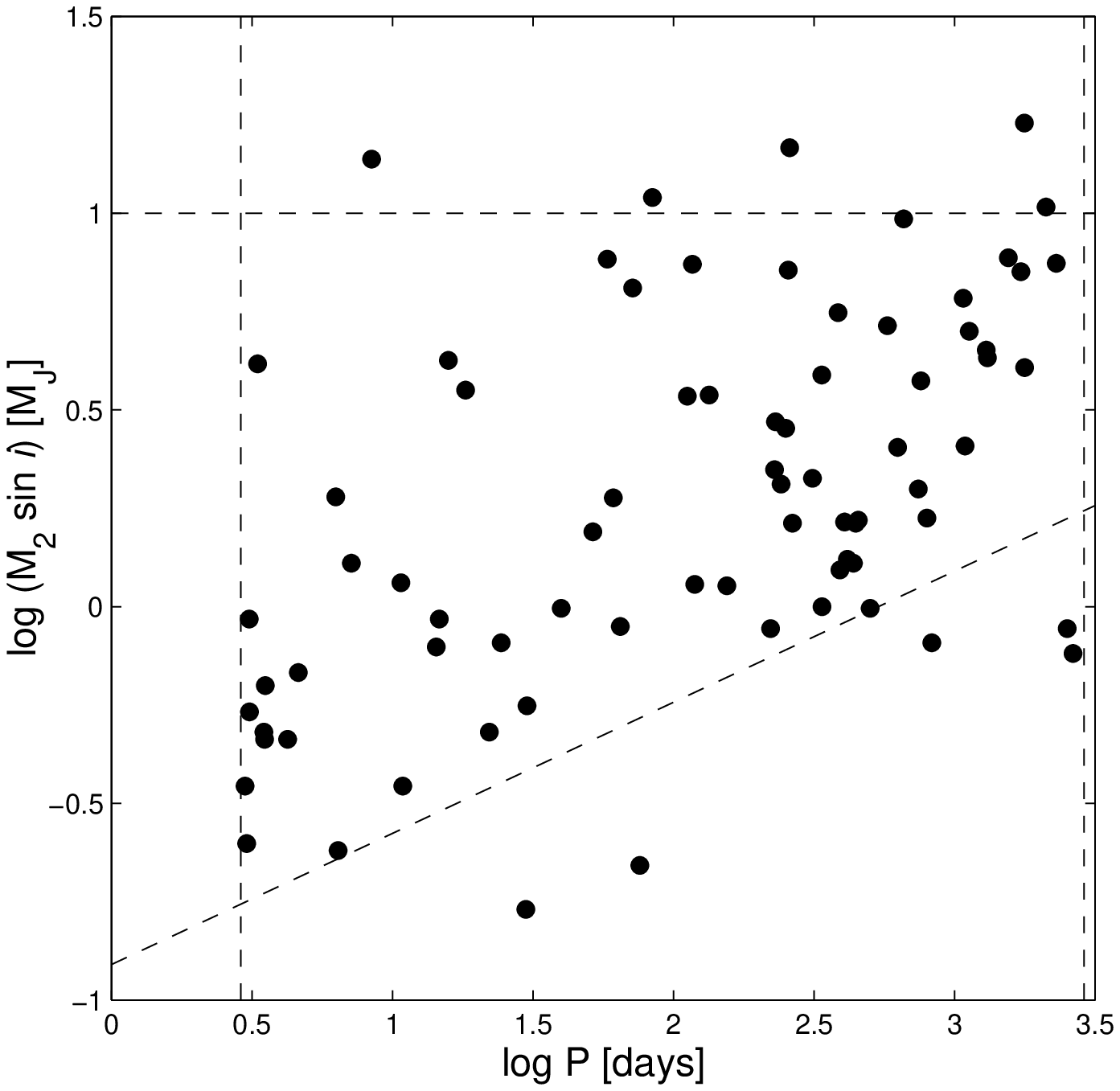}
\caption{The minimum mass vs.\ the period of the extrasolar
planets. The four dashed lines and the trapezoid they form are explained
in the main text. \label{fig1}}
\end{figure}

Figure~\ref{fig1} presents the minimum masses of all known extrasolar
planets as a function of their orbital periods. The data were taken
from the web-site of the California Planet Search Team\footnote{
http://exoplanets.org/planet\_table.txt}, and were updated as of
December 2001. We chose to plot the two axes with logarithmic scales,
because the frequency of planets, up to 10~\MJ, is almost flat in
log~M (Jorissen, Mayor \& Udry 2001; Zucker \& Mazeh 2001; Tabachnik
\& Tremaine 2002), as well as in log~P (Heacox 1999; Stepinski \& 
Black 2001; Mazeh \& Zucker 2002; but see a somewhat different approach
by Tabachnik \& Tremaine 2002).

We concentrated in this analysis on a trapezoidal area in the 
minimum-mass--period parameter space bounded by dashed lines in the figure. The
upper boundary corresponds to the 10~\MJ\ cutoff in planets masses
(e.g., Jorissen, Mayor \& Udry 2001; Zucker \& Mazeh 2001). Although
Zucker \& Mazeh (2001) suggested a probable small higher-mass tail
beyond the 10~\MJ\ line, the distribution is flat up to, probably, 
10~\MJ. We plot the five objects above the line for completeness. The two
vertical lines represent the minimum and maximum orbital periods found
in the sample.

The ascending line at the bottom of the figure corresponds to a constant
radial-velocity amplitude, $K$, of 25\,\ms.  The detection rate of the
present planet-search projects below this line is low. This is easily
seen in the figure, which includes only five planets below this border
line.  Again, we plotted these five planets only for the sake of
completeness. We assume that planets detected above that line have all
been reported.
We now proceed to analyze the 66 planets inside the
trapezoid, assuming a constant detection rate over its area.

A close examination of Figure~\ref{fig1} reveals a paucity
of planets at the high-mass--short-period corner of the trapezoid. Only
three planets appear at that corner. This is certainly not a selection
effect, because planets at that part of the diagram have the largest
radial-velocity amplitudes, and therefore are the easiest to detect.

It is not clear yet what is the shape of the area in which we find low
frequency of planets. It might have, for example, a rectangular shape
bordered by $\log P = 1.6$ and $\log(M_2\sin i)=0.3$, or could have a
wedge shape, bordered by the line from $\left( \log P,\log(M_2\sin i )
\right) = (0.46,0.2)$ to $(1.5,1)$.  In any case, it seems that there
are enough planets in the trapezoid to render this paucity
significant.
 
To estimate quantitatively the statistical significance of the
high-mass--short-period paucity seen in the figure we first consider
the mass-period correlation coefficient of the sample of planets in
the trapezoid. The resulting value was 0.661. 
We claim that this high value means there is a real correlation in
the planets population --- higher than the one induced by the
selection effect. In terms of statistical hypotheses testing, we have
to reject the null hypothesis 
that there is no mass-period correlation in the planets population,
and the correlation we find in the sample comes solely from the
wedge-shape of the area removed by the selection effect. 

\begin{figure}
\plotone{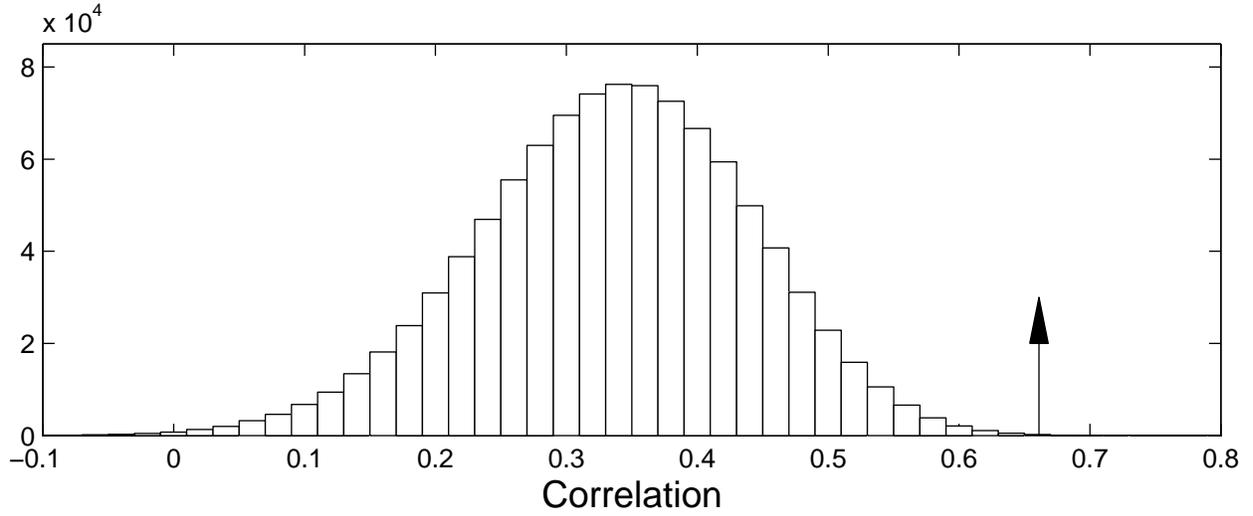}
\caption{Histogram of the correlation coefficients calculated for
  random samples drawn out of a uniform distribution in log-mass and
  log-period. \label{fig2}}
\end{figure}

To assess the statistical significance of the null hypothesis
rejection we used Monte-Carlo simulations in which we created an
artificial sample, randomly drawn out of a two-dimensional uniform
distribution in log-mass and log-period, between the period limits of
the trapezoid, and between 0.175 and 10~\MJ. To simulate the
selection effect we have discarded every planet whose implied radial
velocity was too small and drawn another one instead, until we had in
hand 66 planets.  The process was repeated $10^6$ times, calculating
the correlation for each simulated sample.  Figure~\ref{fig2} shows
the histogram of the simulated correlation coefficients together with
the value derived for the original sample.  The simulated correlation
values distributed around a mean value of 0.336 with a standard
deviation of 0.104.  Only 208 simulations yielded a value larger than
0.661. Thus, we can conclude that the null hypothesis of two
uniform uncorrelated random distributions of log-mass and log-period 
can be rejected with a 99.98\% confidence level. Of course, in the 
currently small sample of planets, even a single additional detection 
of a planet with extremely short period and large mass can alter the
results significantly. 

The seminal work of Tabachnik \& Tremaine (2002) has used maximum
likelihood calculation to study the mass and period distributions of the
extrasolar planets. Assuming the two distributions have power-law
shape and are mutually independent, they derived a positive power of
$0.26\pm 0.06$ for the period distribution and a negative one, 
$-0.12\pm 0.10$, for the
mass distribution. Such a distribution creates a deficiency of
high-mass--short-period planets that might have produced the effect we
report here. In order to check whether this is the case, we repeated
the simulation process with Tabachnik and Tremaine's distribution.
Indeed, the correlation values were somewhat higher than in the
flat distribution case, with an average value of 0.403 and
standard deviation 0.109.  However, here again only 3823 simulations
out of $10^6$ yielded a value larger than 0.661. We can, therefore,
reject this hypothesis with a 99.62\% confidence level. We also
checked the extremely unlikely 
case where the true values of both exponents are 2-$\sigma$ away
from the ones derived by Tabachnik \& Tremaine (2002). Two independent
distributions, with $0.38$
for the exponent of the 
period distribution and $-0.32$ for that of the mass distribution,  
are still rejected at a 98\% confidence level. 
The somewhat lower significance of the last rejection indicates that 
we can not reject all possible hypotheses
where the mass and the period are uncorrelated, and one might come up
with a specific distribution that will reproduce the reported effect,
without a real correlation between the two variables.

\section{Planets in Binary Stars}

Having established the significance of the paucity of the short-period
massive planets, we can examine Figure~\ref{fig1} and try to see what
distinguishes the few planets in the high-mass--short-period corner.
The three planets that appear to be somewhat isolated in that corner
of the figure are $\tau$\,Boo\,b, HD\,195019\,b and Gl\,86\,b --- all
of them are planets found in wide stellar binaries (Hale 1994; Fischer et
al.\ 1999; Els et al.\ 2001).  This raises the possibility that
planets in binary stellar systems have a different mass-period
distribution. If this is true, it can tell us about the possible
effect the binarity of the parent star might have on the formation
(Boss 1998; Nelson 2000) and orbital evolution of their planets (e.g.,
Mazeh, Krymolowski \& Rosenfeld 1997; Holman, Touma \& Tremaine 1997;
Innanen et al.\ 1997; Holman \& Wiegert 1999).

\begin{figure}
\plotone{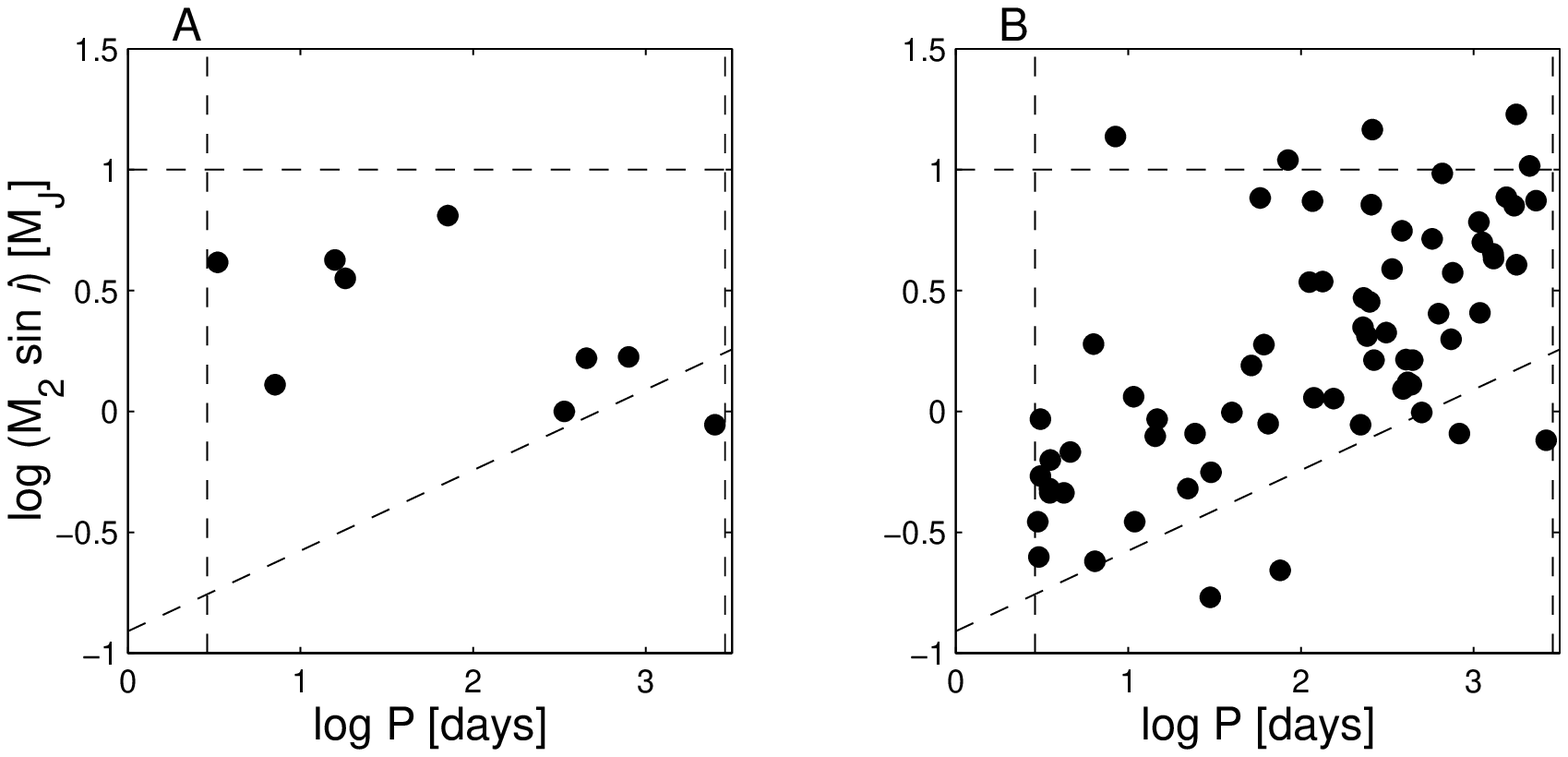}
\caption{The minimum mass vs.\ the period of the extrasolar
planets for the binary stars (A) and the non-binaries (B). The four
dashed lines and the trapezoid they form are explained
in the main text. \label{fig3}}
\end{figure}

In order to check this hypothesis we assembled a subsample of all the
planet-hosting stars that we can safely tag as binaries. Using the WDS
catalog, we decided, somewhat arbitrarily, to consider only those
binaries whose projected separation is smaller than 1000\,AU, assuming
wider binaries would not have influenced the formation and evolution
of their planets. We also discarded binaries that have angular
separations larger than 10\arcsec, whose secondaries were observed to
be fainter than 12th magnitude, and had no other evidence for a
physical association (mainly common proper motion). We added to the
final list Gl\,86 which was announced after the most recent publication
of the WDS. We were left with nine binary stars: HD\,142, Gl\,86,
HD\,19994, $\epsilon$\,Eri, $\tau$\,Boo, HD\,178911\,B,
16\,Cyg, HD\,195019, HD\,217107. Figure~\ref{fig3} shows the two
separate subsamples of planets. $\epsilon$\,Eri lies outside our
nominal trapezoid and thus only eight planets constitute our binary
subsample.

The difference between the two subsamples in Figure~\ref{fig3} is
striking. Apart from the enhanced paucity of planets in the upper left
corner of the figure in the single-star population, the binary
population shows an opposite trend. A {\it negative}-slope
straight line, with a slope of $-$0.15, can be fitted to this
small subsample. This is opposed to the slope of the single-star
planets, for which we fit a straight line with a slope of $+$0.33.
As we showed in the previous section, only part of this positive
slope is due to the specific shape of the trapezoid considered, and
the population of planets does show a positive slope because of the
short-period--high-mass paucity. The remarked difference is also reflected
by the negative mass-period correlation of $-$0.459 for the binary planets.

\begin{figure}
\plotone{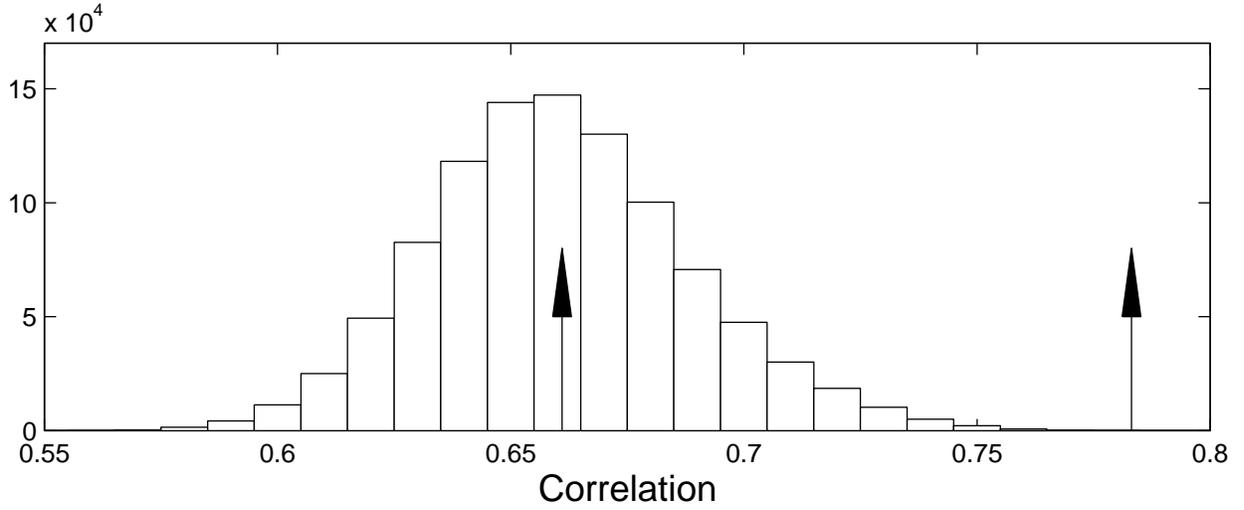}
\caption{Histogram of the correlation coefficients calculated for
  $10^6$ times of randomly removing 8 planets out of the original sample. The
  arrows indicate the values of the correlation before removing the 8
  planets found in binary systems (left arrow) and after removing them
  (right arrow). \label{fig4}}
\end{figure}

The difference between the two populations is manifested also in the
{\it higher} value of the correlation coefficient of the subsample of
58 single-star planets --- 0.783.  In order to test the significance
of the difference between the sample with and without the binary
planets, we ran Monte-Carlo simulations again. Each iteration
consisted of removing a random set of 8 planets from the original
sample of 66 and re-calculating the correlation. The results of $10^6$
iterations are depicted in Figure~\ref{fig4}. The simulated values
average at 0.661, the original value of the parent sample, and have
a standard deviation of 0.028. Only 17 out of $10^6$ simulations
yielded a value higher than 0.783, thus implying a significance of
99.998\% to the higher correlation of the single-star planets.

\section{Discussion}

We have presented evidence for a substantial deficit of massive
planets with short orbital periods. We have shown
that the correlation seen in the sample of
known planets is higher than the one predicted by published
independent power-law models and the selection effect, although we
need more points to corroborate our finding. We regard this
deficit as a refinement of the initial surprising discoveries of giant
planets in close orbits (e.g., Mayor \& Queloz 1995). We have shown
that only planets with masses below about 2~\MJ\ can frequently be
found orbiting single stars with periods shorter than about 40 days.
This may serve to refine or update the models that were devised to
explain the existence of 51\,Peg-like planets.

We can divide the models for the existence of planets with short
orbital periods into three broad categories. The most favored model is
that of planetary migration (e.g., Lin, Bodenheimer, \& Richardson
1996). In this model a planet is formed by core accretion at a
distance of the order of 5~AU or further (but see Bodenheimer,
Hubickyj \& Lissauer 2000 for a somewhat different approach), and then
is pushed toward the parent star by interaction with the accretion
disk. Alternative models are migration by interaction with other
planets (e.g., Weidenschilling \& Marzari 1996) or planetesimals
(Murray et al.\ 1998). A completely different approach assumes
planet formation by disk instability (e.g., Boss 1997a). Obviously, if 
the instability ends up as a planet far away from the parent star, one
needs a migration mechanism to account for the close-in planets (Boss
1997b). In what follows we will try to comment on the implications
of our findings on each of the three categories of models.

Two effects within the migration scenario can contribute to the
paucity of massive planets with small orbits:

(I) Massive planets open a gap in the disc, and consequently slow
their migration rate substantially (e.g., Ward 1997; Trilling et
al.\ 1998; Nelson et al.\ 2000). We expect, therefore, to find the more
massive planets at distances closer to their formation sites.

(II) Trilling et al.\ (1998) pointed out that when planets get too
close to their parent stars they loose substantial fraction of their
mass through Roche-lobe overflow.  In the specific parameters presented by
Trilling et al., planets above 3.4~\MJ\ do not migrate significantly
from their formation site. Most of the planets with initial masses
below 3.4~\MJ\ loose substantial fraction of their mass through
Roche-lobe overflow. In fact, planets with initial masses below
3.36~\MJ\ loose most of their mass, and are left with masses smaller
than 0.4~\Mo.

The two effects, discussed already by theoretical studies of the
migration model, contribute to the paucity of very massive planets
with small orbits. The second effect contributes to the mass-radius
correlation for short distances, on the order of tenths of an AU,
whereas the first effect dominates the correlation for larger
distances. Actually, Trilling et al.\ (1998) already published a
mass-period diagram with a series of models they ran (see their
Figure~7), in which the effect we find here can be clearly seen.

However, the first effect can be canceled out if the planet mass
depends on the disk mass. This is so because the size of the gap
depends on the ratio between the planet mass and the disk mass
(e.g., Trilling et al.\ 1998).  Suppose, for example, that the planet mass
scales with the disk mass. Then more massive planets could move
inwards before they opened a gap as much as less massive planets do.
Therefore, to account for the correlation we see in the data we have
to assume that planets are formed with {\it masses that
do not scale with the disk mass}. In other words, planets with
different masses can be formed in disks with similar masses.

Interaction with other planet(s) was suggested mainly to explain the
high eccentricities observed for some of the known extrasolar planets
(e.g., Weidenschilling \& Marzari 1996, Rasio \& Ford 1996; Ford,
Havlickova \& Rasio 2001). A few models include an accompanying disk
that absorbs the angular momentum necessary to enable the migration
(e.g., Murray, Paskowitz \& Holman 2001).

Suppose the interaction with the other, as yet undetected, planet is
the dominant mechanism for the migration of the known planet.  Such a
scenario can account for the observed mass-period correlation if the
mass of the undetected planet is independent of the mass of
the known planet.  In such a case, the ability of the unseen planet to
push the known planet to smaller radii is limited only to small-mass
planets, consistent with the mass-period correlation we see in
the data. If, on the other hand, the mass {\it ratio} between the planets is
similar in all planetary systems, the migration caused by the
planet-planet interaction should not be limited to small-mass planets,
contrary to our findings.

The same argument applies to migration driven by interaction with
planetesimals. Murray et al.\ (1998) comment that ``if the mass of the
planetesimal disk interior to the planet is of order of the planet
mass, the planet can migrate nearly to the surface of the star''. In
other words, the migration depends on the mass ratio between the
planet and the planetesimal disk. Therefore, to account for the
observed mass-period correlation we have to assume that the planet mass does
not scale with the mass of the planetesimal disk. 

Planet formation via disk instability (Boss 1997a; 1998a; 2000; 2001) could
also account for the mass-period correlation, in principle, if we
assume an in-situ formation for this model.  We can speculate that the
mass of the forming planet depends on the mass available in the disk
at the vicinity of the instability.  Suppose, for example, that the
mass associated with the instability is some fraction of the mass
within the radius of the instability center.  This can lead
to more massive planets forming at large distances, hence the
effect we have detected. However, one still needs to establish by
detailed numerical simulations that this model can work at small
distances from the star. If, on the other hand, the disk instability
can work only at large distances, and thus still requires a
mechanism for migration, then dependence of the planet mass on the
disk mass might make it difficult to account for the paucity we
detected. 

Two theoretical studies have considered the implication of binarity of
a star on the formation of planets around one of its components.  One
report on numerical study (Boss 1998b) indicated that the presence of a
stellar companion can induce a rapid instability even for disks that
are stable otherwise.  A somewhat more recent work (Nelson 2000)
suggested an opposite effect, claiming that planets are unlikely to
form in certain binary systems. Both works dealt only with a companion
at a distance of 40--50 AU, and more comprehensive studies are needed
to explore the implication of a companion on planet formation. In any
case, both works indicate that the planets in binary systems might
not have the same mass-period distribution, consistent with our
findings.

To summarize, it seems as if almost all models for the existence of
close-in giant planets can account for the mass-period correlation of
the single-star planets, although the correlation seems a more natural
outcome of the model that assumes migration driven by a disk-planet
interaction.  In any case, this correlation can put, in
principle, some constraints on the different models.  The number of
planets known today is only marginally enough to characterize the
details of the short-period--massive-planets deficit, apart from
establishing its existence. Nevertheless, we already can conclude that
for all models we need planet masses that do not scale with the mass
of the disk/planetesimals/other planets. More data can illuminate the
finer details of this phenomenon and help to better tune the theories
for close-in giant planets.

\begin{acknowledgements}

We wish to thank S.\ Tremaine for important comments on an earlier
version of the manuscript.  We are indebted to the referee, M. Holman, 
for his
careful reading of the paper, his useful comments and advise, and for
pointing out an error in one of our simulations. This research has
made use of the Washington Double Star Catalog maintained at the
U.S. Naval Observatory, and was supported by the Israeli Science
Foundation (grant no. 40/00).

\end{acknowledgements}

\end{document}